\documentclass[10pt,conference]{IEEEtran}

\usepackage{amsmath,amssymb,amsfonts,amsthm}
\usepackage{graphicx}
\usepackage{xcolor}
\usepackage{booktabs}
\usepackage{multirow}
\usepackage{algorithm}
\usepackage{algorithmic}
\usepackage{cite}
\usepackage[hidelinks]{hyperref}
\usepackage{physics}
\usepackage{balance}

\usepackage{subcaption}

\title{Resource Analysis for Quantum Simulation of Spatially Varying Transport–Reaction Equations
\vspace{-0.0em}
{\normalfont\footnotesize\itshape
\parbox{\textwidth}{
\centering
Preprint. Accepted for publication in the proceedings of
QC4PDE 2026: Quantum Computing for PDEs---Algorithms, Applications \& Challenges,
IEEE Quantum Week (QCE) 2026.
}}
}

\author{
\IEEEauthorblockN{
Benjamin Jasperson\IEEEauthorrefmark{1},
Krishna Garikipati\IEEEauthorrefmark{1},
Siddhartha Srivastava\IEEEauthorrefmark{2}\IEEEauthorrefmark{3}
}
\IEEEauthorblockA{\IEEEauthorrefmark{1}
Department of Aerospace and Mechanical Engineering, Los Angeles, USA\\
Email: \{bjaspers, garikipa\}@usc.edu}
\IEEEauthorblockA{\IEEEauthorrefmark{2}
Department of Aerospace Engineering, Auburn University, Auburn, USA\\
Email: sis0024@auburn.edu}
\IEEEauthorblockA{\IEEEauthorrefmark{3}
Corresponding author
}
}
\begin{document}
\maketitle

\begin{abstract}
Spatially varying coefficients are the primary source of circuit complexity in quantum simulation of linear advection-diffusion–reaction equations. This work presents a resource analysis of single-step quantum propagators constructed using sparse FABLE block encodings, Quantum Singular Value Transformation, and linear-combination-of-unitaries. We derive theoretical estimates for qubit count, gate complexity, and circuit depth in terms of the spatial discretization, sparsity, and polynomial degree, and compare these predictions with synthesized quantum circuits for one- and two-dimensional variable-coefficient problems. The resulting analysis quantifies the cost of encoding realistic transport operators and provides practical resource estimates for near-term implementations.
\end{abstract}



\section{Introduction}

Partial differential equations (PDEs) describing transport, diffusion, and reaction processes arise in numerous scientific and engineering applications, including heat and mass transfer, reactive transport, biological pattern formation, and population dynamics. Numerical simulation of these systems often requires the solution of large-scale sparse linear systems or repeated time integration, making them attractive candidates for quantum algorithms capable of exploiting high-dimensional Hilbert spaces.

Quantum algorithms for PDEs have developed along several complementary directions. Variational quantum algorithms have been developed for solving PDEs through parameterized quantum circuits and residual minimization~\cite{Yuan2019}. Meanwhile, quantum annealing has been explored for energy-based formulations of differential equations~\cite{Srivastava_2019}. Another active direction has focused on translating established numerical discretizations, such as finite element, finite difference, and spectral methods, into quantum computational frameworks, thereby enabling the application of quantum algorithms to sparse systems arising from PDE discretization~\cite{Montanaro2016,Deiml2025,Alkadri2025,Helle2026,febrianto2026quantum}. These developments naturally motivated the application of quantum linear system algorithms, beginning with the Harrow–Hassidim–Lloyd (HHL) algorithm \cite{harrow2009quantum} and its subsequent extensions, for solving the large sparse linear systems arising from PDE discretizations~\cite{gopalakrishnan2024solving}. More recently, Quantum Singular Value Transformation (QSVT) \cite{gilyen2019quantum} has emerged as a powerful framework for implementing general matrix functions through block encodings, enabling efficient approximation of matrix exponentials and other operator functions central to time-dependent PDE simulation~\cite{Helle2026,Novikau2025}. 

In this work, we consider a transport--reaction equation

\begin{equation}
\frac{\partial c}{\partial t}
=
\nabla\cdot\!\left(D(\mathbf{x})\nabla c\right)
-
\nabla\cdot\!\left(\mathbf{v}(\mathbf{x})c\right)
+
r(\mathbf{x})c,
\label{eq:adr}
\end{equation}

defined on the periodic domain $\Omega=[0,L]^d$ with periodic boundary conditions. The coefficient fields $D(\mathbf{x})$, $\mathbf{v}(\mathbf{x})$, and $r(\mathbf{x})$ are assumed to be smooth and periodic, and denote the diffusivity, advection velocity, and reaction rate, respectively.

Sparse discretizations of \eqref{eq:adr} naturally lead to sparse matrices, however, the presence of spatially varying coefficients substantially increases the complexity of quantum implementations. Unlike constant-coefficient operators, where the matrix structure is highly regular, variable-coefficient transport operators require encoding a large number of nonuniform matrix entries while preserving sparsity. Consequently, for practical PDE simulations, the dominant computational cost often lies in constructing efficient block encodings~\cite{Camps2023,FABLE}, whose resource requirements are incurred repeatedly during the polynomial transformations implemented by QSVT.

The objective of this paper is to quantify the computational resources required for quantum simulation of realistic variable-coefficient transport--reaction equations. We consider a single-step exponential propagator for the semi-discrete system, following the QSVT-based framework for linear PDE simulation (c.f.~\cite{Helle2026}). The next section outlines the construction of the corresponding quantum circuit, while the subsequent sections present asymptotic estimates for qubit count, gate complexity, and circuit depth, highlighting their dependence on the spatial discretization, sparsity, and polynomial degree. Finally, we validate these theoretical predictions through numerical resource analysis of one- and two-dimensional transport--reaction problems using quantum circuits synthesized with the PennyLane software framework~\cite{bergholm2018pennylane}.

\section{Methodology}

A second-order finite-difference discretization of \eqref{eq:adr} on a periodic Cartesian grid with
$N_g=2^n$ degrees of freedom in each coordinate direction produces a semi-discrete system with
$N=N_g^d=2^{nd}$ unknowns,
\begin{equation}
\frac{d}{dt}\ket{c(t)}=A\ket{c(t)},
\end{equation}
where $A\in\mathbb C^{N\times N}$ is a sparse, generally non-Hermitian matrix. The solution over one timestep $\Delta t$ is
$
\ket{c_{m+1}}
=
e^{\Delta tA}\ket{c_m}.
$

Following the Cartesian decomposition, we write
$
A=A_S-iA_H,
$
where
$
A_S=(A+A^\dagger)/2
$
and
$
A_H=i(A-A^\dagger)/2
$
are Hermitian. The operator $A_S$ governs the non-unitary (amplitude-changing) component of the evolution, while $A_H$ generates the unitary component. The exponential propagator is approximated using the first-order Lie product
\begin{equation}
e^{\Delta tA}
=
e^{\Delta t(A_S-iA_H)}
\approx
e^{-i\Delta tA_H}
e^{\Delta tA_S},
\label{eq:split}
\end{equation}
whose local error satisfies
$
e^{\Delta t(A_S-iA_H)}
-
e^{-i\Delta tA_H}
e^{\Delta tA_S}
=
\mathcal O\!\left(\Delta t^2\|[A_S,A_H]\|\right).
$
For spatially varying transport and reaction coefficients, $A_S$ and $A_H$ generally do not commute, making the splitting error {\color{black} locally second-order and globally first-order in the timestep.

\smallskip
\noindent\textbf{Regularizing the operators.}
QSVT requires the block-encoded operator to have singular values contained in $[-1,1]$. Ideally, the operators would be scaled using their spectral norms; however, estimating spectral norms or extremal eigenvalues becomes increasingly expensive for the large sparse matrices arising from PDE discretizations. Instead, we employ inexpensive bounds derived from Gershgorin's theorem, which require only row-wise operations and are therefore naturally parallelizable.

The Hermitian operator $A_S$ is first shifted according to
$
\tilde A_S=A_S-\mu_GI,
$
where
$
\mu_G=\max_i\!\left((A_S)_{ii}+\sum_{j\neq i}|(A_S)_{ij}|\right)
$
is a Gershgorin upper bound on the largest eigenvalue of $A_S$. Consequently,
$
\lambda(\tilde A_S)\le0,
$
ensuring that $e^{t \tilde A_S}$ is contractive for $t\geq 0$. The shifted operator is then normalized using the row-sum bound
$
\alpha_S=\|\tilde A_S\|_\infty,
$
while the Hamiltonian component is scaled by
$
\alpha_H=\|A_H\|_\infty.
$
Defining
$
\bar A_S=\tilde A_S/\alpha_S,
$
$
\bar A_H=A_H/\alpha_H,
$
$
\tau_S=\alpha_S\Delta t,
$
and
$
\tau_H=\alpha_H\Delta t,
$
the timestep propagator becomes
\begin{equation}
e^{\Delta tA}
\approx
e^{\mu_G\Delta t}
e^{-i\tau_H\bar A_H}
e^{\tau_S\bar A_S}.
\label{eq:scaled_split}
\end{equation}

\noindent\textbf{Sparse block encoding.}
The scaled operators are represented using Sparse-FABLE block encodings~\cite{Camps2023,FABLE}. A unitary $U_M$ is a block encoding of a matrix $M$ if
$
(\langle0|^{\otimes a}\!\otimes I)
U_M
(|0\rangle^{\otimes a}\!\otimes I)
\approx
M/s,
$
where \(s\) denotes the block-encoding normalization.
For structured finite-difference discretizations, the sparse matrix admits the decomposition
$
M=\sum_{\mathbf{s}\in\mathcal S} D_{\mathbf{s}}P_{\mathbf{s}},
$
where $D_{\mathbf{s}}$ are diagonal matrices containing the spatially varying coefficients associated with stencil offset $\mathbf{s}$. $P_{\mathbf s}$ are permutation matrices that translate the solution vector by the stencil offset $\mathbf s$ under periodic boundary conditions. Sparse-FABLE encodes the stencil offsets in a sparse selector register while the diagonal coefficient matrices are implemented through multiplexed rotations, yielding a block encoding whose width grows logarithmically with the number of stencil offsets $|\mathcal S|$. The selector register prepares a uniform superposition over the padded stencil labels. Consequently, the block encoding normalization for each scaled operator is $s=2^{\lceil\log_2|\mathcal S|\rceil}$. The QSVT polynomials are synthesized using the rescaled evolution parameter
$\bar\tau=s\tau$, ensuring that
\begin{equation}
P_{\bar\tau}(M/s)\approx e^{\tau M},
\end{equation}
where $P_{\bar\tau}$ denotes the synthesized QSVT  polynomial evaluated on the encoded operator.

For example, a one-dimensional second-order stencil has
$\mathcal S_S=\{-1,0,+1\},
\mathcal S_H=\{-1,+1\},
$ while a two-dimensional five-point stencil corresponds to
$ \mathcal S=\{(0,0),(\pm1,0),(0,\pm1)\}. $
Higher-order and higher-dimensional discretizations follow analogously.

\smallskip
\noindent\textbf{Exponential operator construction.}
Following Ref.~\cite{Helle2026}, the exponential propagators are implemented using QSVT. Since QSVT synthesizes bounded polynomial transformations of fixed parity, each exponential is decomposed into even and odd components. For the symmetric branch,
\begin{equation}
e^{\tau_Sx}
=
\cosh(\tau_Sx)
+
\sinh(\tau_Sx).
\end{equation}
Because the hyperbolic functions generally exceed unity on $[-1,1]$, QSVT is applied to the normalized polynomial approximations
\begin{equation}
P_{\rm even}(x)
\approx
\frac{\cosh(\bar\tau_Sx)}
{\cosh(\bar\tau_S)},
\qquad
P_{\rm odd}(x)
\approx
\frac{\sinh(\bar\tau_Sx)}
{\sinh(\bar\tau_S)},
\end{equation}
where $\bar\tau_S=s_S\tau_S$. The exponential is reconstructed through the two-term Linear Combination of Unitaries (LCU)
\begin{equation}
e^{\tau_Sx}
\approx
\cosh(\bar\tau_S)P_{\rm even}(x)
+
\sinh(\bar\tau_S)P_{\rm odd}(x).
\end{equation}

The Hamiltonian branch is treated analogously using
\begin{equation}
e^{-i\tau_Hx}
=
\cos(\tau_Hx)
-
i\sin(\tau_Hx),
\end{equation}
with independent QSVT approximations of the truncated Taylor series expansions of cosine and sine functions synthesized using
$\bar\tau_H=s_H\tau_H$.

Finite-degree QSVT implements a unitary signal transformation whose selected block generally has the form
$
P(x)+iQ(x),
$
where $Q(x)$ is the complementary polynomial required by unitarity and is not necessarily small. The desired real polynomial is therefore extracted through the Hermitian construction
$
\mathcal R(U_P)
=
\frac{1}{2}
\left(
U_P
+
U_P^\dagger
\right),
$
implemented coherently using an additional selector qubit. The symmetric and Hamiltonian propagators are finally assembled through successive two-term LCUs, yielding a quantum circuit that implements the propagator in~\eqref{eq:scaled_split}.

\section{Results}

\subsection{Theoretical resource estimates}
\label{sec:theor_resources}
The computational cost of the propagator is determined by the block encodings, QSVT polynomial degree, and LCU constructions introduced in the previous section. We estimate the required qubits, gate complexity, approximation error, and postselection probability under the assumption of exact block encodings, and subsequently discuss the effect of approximate encodings.

\smallskip
\noindent\textbf{Circuit width.}
Consider a $d$-dimensional Cartesian discretization with $N_g=2^n$ grid points per coordinate direction, giving
$
N=N_g^d=2^{nd}
$
degrees of freedom. A second-order finite-difference stencil requires $2d+1$ stencil offsets for the symmetric operator and $2d$ offsets for the Hamiltonian operator. The work register therefore contains $nd$ qubits, while the corresponding Sparse-FABLE selector registers require
$
\lceil\log_2(2d+1)\rceil
$
and
$
\lceil\log_2(2d)\rceil
$
qubits, respectively. The present implementation further employs six auxiliary qubits: two FABLE amplitude ancillas, two selector qubits for the Hermitian real-part constructions, and two selector qubits for the outer LCU combinations. The total qubit count for one timestep is therefore
\begin{equation}
n_{\rm qubits}
=
nd
+
\left\lceil\log_2(2d+1)\right\rceil
+
\left\lceil\log_2(2d)\right\rceil
+
6.
\end{equation}
Thus, while the auxiliary circuitry contributes a fixed overhead, the total qubit count grows linearly with the number of spatial qubits.

\smallskip
\noindent\textbf{Circuit depth and gate count.}
We estimate circuit size in the native operations used by the construction rather than after decomposition into a minimal universal gate set. In particular, the Sparse-FABLE block encodings use multiplexed $R_y$ rotations and CNOT/controlled-X operations to encode the diagonal value table and implement the stencil shifts. For the Hamiltonian branch, which contains complex-valued entries, an additional controlled $R_z$ phase oracle is used. Since the circuit is laid out serially, the operation count also provides a useful proxy for circuit depth before hardware-specific compilation.

Let $G_S$ and $G_H$ denote the native-operation counts for one Sparse-FABLE block encoding of $\bar A_S$ and $\bar A_H$, respectively. For a second-order stencil, these scale as
\begin{equation}
G_S=\mathcal O\!\left((2d+1)N\right),
\qquad
G_H=\mathcal O\!\left((2d)N\right),
\end{equation}
up to the cost of the controlled shift networks and, for $G_H$, the controlled phase oracle. A QSVT sequence of degree $r$ invokes the block encoding $\mathcal O(r)$ times. In the present construction, each exponential branch is split into two parity components, and each component is passed through the real-part construction
$
\mathcal R(U_P)=\frac12(U_P+U_P^\dagger),
$
which doubles the number of QSVT calls. Assuming the same QSVT degree $r$ for all four functions, one timestep uses approximately
$
4r\,G_S+4r\,G_H
$
block-encoding calls, plus lower-order projector-phase, selector, and LCU operations. Therefore,
\begin{equation}
G_{\rm step}
=
\mathcal O\!\left(
r(G_S+G_H)
\right)
=
\mathcal O\!\left(
r(4d+1)N
\right).
\end{equation}

\smallskip
\noindent\textbf{Approximation error.}
Assuming exact Sparse-FABLE block encodings, the one-step approximation error is governed by two independent sources: operator splitting and QSVT polynomial approximation. By the triangle inequality,
\begin{equation}
\|U_{\rm exact}-U_{\rm impl}\|
\le
\|U_{\rm exact}-U_{\rm split}\|
+
\|U_{\rm split}-U_{\rm impl}\|,
\end{equation}
where
$
U_{\rm exact}=e^{\Delta tA},
$
$
U_{\rm split}
=
e^{-i\tau_H\bar A_H}
e^{\tau_S\bar A_S},
$
and
$
U_{\rm impl}
$
denotes the implemented split propagator. 

The first term is the Lie-splitting error, which vanishes only when the symmetric and Hamiltonian operators commute.
\begin{equation}
\|U_{\rm exact}-U_{\rm split}\|
=
\mathcal O\!\left(
\Delta t^2
\|[A_S,A_H]\|
\right),
\end{equation}

The second term arises from the polynomial approximation of the exponential functions. Since the target functions are analytic on the bounded interval $[-1,1]$, the QSVT approximation converges exponentially with polynomial degree,
\begin{equation}
\|U_{\rm split}-U_{\rm impl}\|
=
\mathcal O
\!\left(
e^{-c(r-\bar\tau)}
\right),
\end{equation}
where
$
\bar\tau=\max(\bar\tau_S,\bar\tau_H)
$
and $c>0$ is a constant determined by the polynomial approximation.
Consequently,
\begin{equation}
{
\|U_{\rm exact}-U_{\rm impl}\|
=
\mathcal O\!\left(
\Delta t^2
\|[A_S,A_H]\|
+
e^{-c(r-\bar\tau)}
\right).
}
\end{equation}

The block encodings used throughout this work are exact up to numerical precision. Consequently, no additional block-encoding error appears in the theoretical estimates. If approximate block encodings are employed, the corresponding implementation error is repeatedly queried by QSVT and typically scales linearly with the polynomial degree r. Hence, it will contribute an additional
$
\mathcal O(r\varepsilon_{\rm enc})
$
term to the total error.

\smallskip
\noindent\textbf{Postselection probability.}
The symmetric and Hamiltonian propagators are implemented through successive postselected QSVT circuits, so the success probability of one timestep is
\begin{equation}
p_{\rm succ}(\psi)
=
\frac{
\left\|
e^{\tau_S\bar A_S}
|\psi\rangle
\right\|^2
}{
\left(\cos(\bar\tau_H)+\sin(\bar\tau_H)\right)^2
\left(\cosh(\bar\tau_S)+\sinh(\bar\tau_S)\right)^2
},
\end{equation}
where the unitary Hamiltonian propagator preserves the state norm, and the denominator consists of the normalization constants introduced by the two-term LCU reconstructions of the hyperbolic and trigonometric branches. 

\textit{Remark:}
The expressions above assume the ideal polynomial normalizations
$\cosh(\bar\tau_S)$,
$\sinh(\bar\tau_S)$,
$\cos(\bar\tau_H)$,
and
$\sin(\bar\tau_H)$.
In practice, the polynomial supplied to the QSVT phase-synthesis routine may require an additional normalization factor $\gamma\ge1$ to satisfy the boundedness constraint $|P(x)|\le1$ on $[-1,1]$. Here, $\gamma$ denotes the smallest such scaling factor. The corresponding LCU normalization is multiplied by $\gamma$, reducing the postselection probability by a factor of $1/\gamma^2$ for each branch.

\subsection{Numerical Results}
To validate our theoretical estimates, we evaluate one- and two-dimensional transport--reaction equations 
using circuits synthesized via PennyLane~\cite{bergholm2018pennylane}. Figure~\ref{fig:placeholder} illustrates the compiled native gate count per timestep versus problem size $N = 2^{nd}$.
Here, spatial refinement is coupled with quadratic temporal refinement ($\Delta t \propto \Delta x^{-2}$) to maintain classical accuracy.
The empirical sizes closely mirror the theoretical scaling models established in Section~\ref{sec:theor_resources}. 
This strict first-order scaling across all native gate types confirms that exact Sparse-FABLE block encodings of spatially varying fields dominate the circuit depth overhead. 
Note that these resource counts reflect native circuits prior to decomposition into a minimal universal gate set (e.g., via the Solovay–Kitaev theorem).

\begin{figure}
    \centering
    \includegraphics[width=0.8\columnwidth]{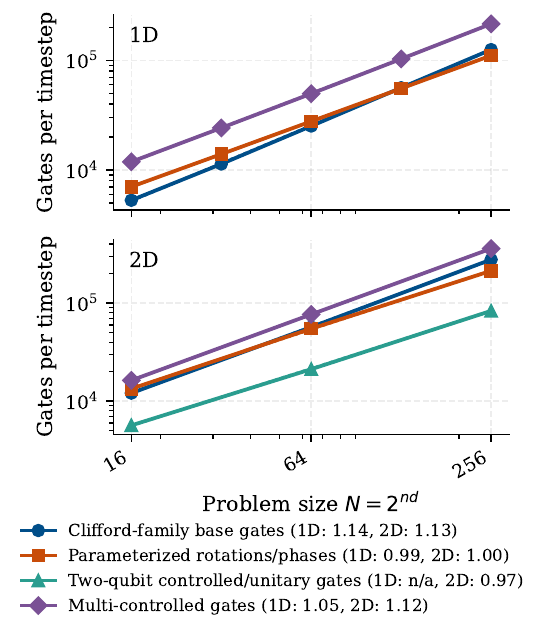}
    \caption{Breakdown of single step gate count type (w/ slope) for 1D and 2D ADR simulations. ($\Delta t \propto \Delta x^{-2}$)}
    \label{fig:placeholder}
\end{figure}


\begin{figure}
    \centering
    \includegraphics[width=0.8\columnwidth]{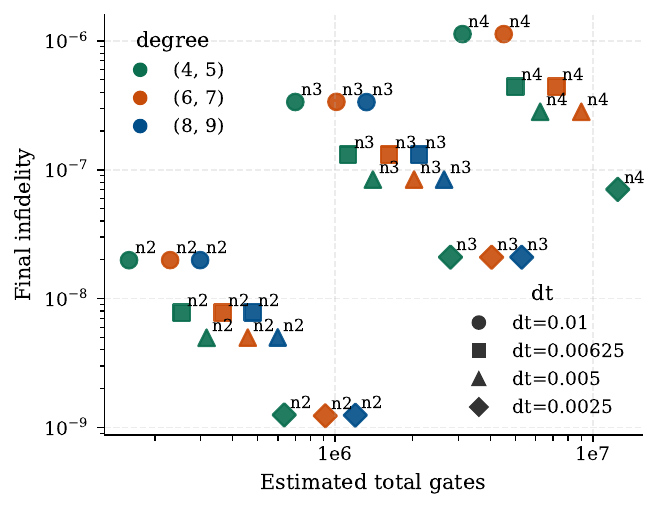}
    \caption{
    Influence of polynomial degree and spatiotemporal resolution on final state infidelity for a 2D system at $T= 0.1$s. 
    Final infidelity is \(1 - |\langle \hat{u}_{\mathrm{DNS}}(T), \hat{u}_{\mathrm{impl}}(T) \rangle|^2\), where \(\hat{u}=u/\|u\|_2\) and \(u_{\mathrm{impl}}(T)\) is obtained by repeatedly applying \(U_{\mathrm{impl}}\).
    }
    \label{fig:cost_vs_final_infidelity}
    \vspace{-1em}
\end{figure}

We then proceed to evaluate the influence of polynomial degree and spatiotemporal resolution on the final state infidelity for the 2D ADR system (Fig.~\ref{fig:cost_vs_final_infidelity})
To retain the temporal accuracy typical of classical methods, refinement of the spatial resolution requires a corresponding quadratic temporal refinement ($\Delta t \propto \Delta x^{-2}$). 
Notably, the results demonstrate that polynomial approximation error remains orders of magnitude below the Lie-splitting error, even at the lowest QSVT degrees evaluated. 
Consequently, increasing the polynomial degree merely shifts the computational cost higher (requiring more gates) without yielding any meaningful reduction in the final state infidelity.

\section{Discussion and Conclusions}

This work provides a practical assessment of the quantum resources required to simulate spatially varying transport--reaction equations using exact block encodings. 
Both the theoretical estimates and compiled PennyLane circuits show that the primary bottleneck is not the number of qubits, but the circuit depth required to exactly embed variable-coefficient operators. 
Although the work-register size grows only linearly with the spatial dimension and logarithmically with the spatial resolution, maintaining practical solution accuracies of under $10^{-6}$ requires compiled circuits containing $10^5$--$10^6$ native gates even for one- and two-dimensional benchmark problems considered here. 
These results suggest that practical quantum simulation of PDEs will depend critically on approximate block encodings that trade controlled approximation error for reduced circuit depth. 
Equally important is the offline classical preprocessing, where Gray-code/M\"ott\"onen synthesis of the Sparse-FABLE value oracle dominates the cost, whereas QSVT phase synthesis depends only on the polynomial degree. 
Together, these results establish a practical resource baseline for future hardware, compiler, and approximate encoding developments for quantum PDE simulation.

\bibliographystyle{ieeetr}
\bibliography{references}

\end{document}